
\magnification=1200
\baselineskip 16pt
\nopagenumbers
\def\RR{{{I\negthinspace\!R}}}
\centerline{\bf The Topological Entropy of One-Dimensional Maps:}
\centerline{\bf Approximations and Bounds}
\bigskip
\centerline {N.J. Balmforth and E.A. Spiegel}
\smallskip
\centerline {\it Department of Astronomy, Columbia University, New York, New
York, 10027}
\medskip
\centerline {C. Tresser}
\smallskip
\centerline {\it IBM T. J. Watson Laboratories, Yorktown Heights, New
York, 10598} 
\medskip
\bigskip
\bigskip
\hrule
\medskip

We present a method for computing the topological
entropy of one-dimensional maps.
As an approximation scheme, the algorithm converges rapidly and
provides both upper and lower bounds.

\medskip

\hrule

\bigskip
\bigskip


\vfill
\eject
\pageno=2

\footline={\hss--\folio\--\hss}
The topological entropy of a dynamical system was introduced in the
sixties as a quantity that is invariant under continuous changes of
coordinates.$^1$  It has better continuity properties than other
indicators of the degree of chaos, the Lyapunov exponents
on which it is an upper bound (supremum).
Further, the topological entropy is well suited to the
theoretical description of the transition to chaos.  For many
families of maps, the transition to complex behavior is reflected
in the topological entropy as a second-order phase transition$^2$
(see figures 1 and 2).  Thus the
topological entropy has figured increasingly in quantifications
of chaos and discussions of the physical effects of chaos. In dynamo
theory, for example, it has been shown that the topological entropy of a
fluid flow is an upper bound on the rate of growth of the magnetic
energy.$^3$

The uses of the topological entropy, especially to measure
the degree of chaos, provide an incentive for developing
good algorithms for calculating it.  We present here a usable, effective
scheme for estimating and bounding the topological entropy, $h$, of a
piecewise monotonic map of the unit interval.
In comparison to known schemes,$^{4,5,6}$
the method is more flexible in
treating maps with multiple turning points,
with at least as much precision.
Our method works for circle maps and can be adapted for maps on graphs.
Maps with plateaus can also be treated as discussed here, but we exclude
them for brevity.

Points of the interval for which a map is
an extremum are called turning points,
including the end points of the interval considered,
and piecewise monotonic maps have a finite number of
these.  The locations of neighboring turning points divide the interval into
segments called {\it laps}. For a piecewise monotonic map, $f$, the
topological entropy is given by the exponential rate of increase with $n$
of the monotonic upper bounds of
any of these quantities:
(a) the number of turning points of $f^n$, (b) the length of the
graph of $f^n$, and (c) the number of periodic orbits of $f^n$.
For such maps,
these properties are equivalent$^7$ to the standard definitions.$^8$

The paradigm of piecewise monotonic maps is the family of sawtooth maps,
whose slopes have constant magnitude, $s$, everywhere.  An important
special case of these is the family of tent maps
$$
f_s(x) = \left\{  \matrix{  s(x-1)+2 &
{\rm for} \quad 0 \leq x < (s-1)/s \cr
                              s(1-x) &
{\rm for} \quad (s-1)/s \leq x \leq 1 .\cr}
\right.
\eqno(1)
$$
According to property (b), the entropy of any sawtooth map
with slopes $\pm s$ is simply
$\log s$.

The central object in our method is the {\sl topological transfer
matrix}, ${\bf M}$.  To define it, we partition
the unit interval, $I$, by an ordered set of points $(x_0, x_1, ...,
x_{k+1})$ so that $I=[x_0,x_{k+1}]$
is subdivided into subintervals $I_i = [x_i,x_{i+1}]$.
While the subintervals $I_i$ are not necessarily laps,
some of them may contain more than one lap. We further partition
such subintervals by the turning points they
contain, and we let $I^{\ell}_i$ be the $\ell^{th}$ lap contained
in $I_i$.  The elements of ${\bf M}$ are then defined as
$$
m_{ij}= \sum_{\,\ell} {|f(I^{\ell}_i)\cap I_j|\over|I_j|},
\eqno(2)
$$
where $|I_i|$ is the length of the interval $I_i$.
Thus $m_{ij}$ is the fractional amount by which $f(I_i)$,
the image of $I_i$, covers the interval $I_j$, giving suitable
weight to any portion that is multiply covered.

In the case of sawtooth maps on the interval $[0,1]$
with local slopes $\pm s$, it is readily verified
that the matrix ${\bf M} = (m_{ij})$ has the eigenvalue $s$, the antilog of
the topological entropy, with corresponding
eigenvector $(|I_0|, |I_1|, .... |I_k|)^{\bf
T}$.  It follows directly from the Perron-Frobenius theorem$^9$ that this is
the maximum eigenvalue.

The formula for $m_{ij}$ is complicated because we allow for
multiple covering of the target interval.  We may avoid this complication
by including all of the turning points of $f$ in the partition
when their number is not too large.  An effective way
subsequently to refine the partition to improve the accuracy is
suggested by another class of maps
for which the topological entropy is known exactly.  These are the
Markov maps, in which the orbit of every turning point is finite.

This time, we choose the partition by forming an
ordered set of points,
$[x_0,...,x_{k+1}]$, from the turning points and their orbits.
Then, for each pair of subintervals,
$(I_i,I_j)$, it can be seen on constructing the partition that
either $f(I_i)$ covers $I_i$ completely or the intersection
of $I_i$ with $f(I_i)$ is empty.  Hence the matrix ${\bf M}$, defined as
before, has only $0$s and $1$s as entries. In this case, it is known that
the entropy for the Markov map is given by the logarithm of the maximal
eigenvalue of ${\bf M}$.$^{10}$

For both the general Markov maps with suitable
partitions and for sawtooth maps, the largest eigenvalue of ${\bf M}$ is
the antilog of the topological entropy.
These results highlight the significance of $\bf M$.
We are lead to construct the
topological transfer matrix for general piecewise monotonic maps
by specifying suitable partitions of $I$.  Several are possible, but one
modeled after the Markov example combines accuracy with ease
of construction.

Consider a piecewise monotonic map, $f$, whose turning points lie
at the locations $y_p$, $p=1,2,...,P$.  For our partition of the
interval, we now make up the ordered set of points $x_i$ from
the turning points and their images, $f^q(y_p)$, $q=0,1,2,...,Q(p)-1$,
counting coincident points only once.
Now, at any level of refinement,
we have $k=\sum_p Q(p)$ points in our partition.  As we iterate each
turning point in such a way that $\min_p Q(p)$ increases, the partition
approximates better and better to a Markov partition.
In the examples discussed below,
we consider the invariant pieces of
the map, and so, as in the Markov case, the limits of the
interval belong to the orbits of other turning points. Hence
the end points need not be treated as distinct turning points.
In such cases, we take $Q$ to be
independent of $p$.  The number of partitions is then given by $k = PQ$
where we further write $N=k-1$ and speak of the $N^{th}$ level of
approximation.  In the exceptional case where a turning
point has a finite orbit we consider only the intervals of finite
length.

As in (2), we introduce the fraction by which the map of $I_i$ covers
$I_j$ and compose the matrix with entries, $m_{ij}$.  Once again,
the entries in the matrix ${\bf M}$ are mostly $0$s and
$1$s, but some columns may contain subcolumns of the form
$\left( a_{i,j}, a_{i,j+1}, ... , a_{i,j+m}\right)^{\bf T}$
where $a_{i,\ell}\ge 0$ and $\sum_{k=0}^m a_{i,j+k}\le 1$.
The first key result is that the log$^+$ of the largest eigenvalue of ${\bf
M}$ is an approximation to the topological entropy
with an error that decreases with increasing $N$,
where log$^+ X$ means the
larger of zero and log$X$.

More importantly,
it is also possible to put bounds on the topological entropy with this
partition. To see this, we associate to the subcolumn $\left( a_{i,j},
a_{i,j+1}, ... , a_{i,j+m}\right)^{\bf T}$ of $\bf M$ of length $m$, the
$m$-simplex, that is, the set of points points $(X_0,...,X_m)$ in
$\RR^{m+1}$  with  $X_n\ge 0$ and $\sum X_n = 1$.  The extremities of this
simplex are at the points $P_0 = \left(1, 0, 0, ..., 0\right)^{\bf T}$,
$P_1 = \left(0, 1, 0, ..., 0\right)^{\bf T},...$, $P_m = \left(0, 0, 0,
..., 1\right)^{\bf T}$.  For $m=0$, we set $P_0 = 0$ and $P_1 = 1$.

We generate a set of matrices by replacing each subcolumn $\left( a_{i,j},
a_{i,j+1}, ... , a_{i,j+m}\right)^{\bf T}$ in $\bf M$ from a given lap by
every possible $P_n$ of the appropriate length, in all combinations.
If we compute the largest eigenvalues of all the matrices so obtained, we
get a collection of numbers whose extrema give lower and upper bounds for
the largest eigenvalue of ${\bf M}$, which follows in consequence of the
Perron-Frobenius theorem.  As can be shown from the kneading theory of
Milnor and Thurston$^{11}$, these extrema give upper and lower bounds on
the topological entropy, and these bounds improve when we use longer
orbits of the turning points for our partitions.  In certain examples,
partitions may contain cells of zero length and a consistent criterion for
computing $m_{ij}$ may be needed to allow for this.

By way of illustration, we show in Fig. 1 the calculation of the
entropy exponent $\lambda$, the greatest eigenvalue of ${\bf M}$, for the
quadratic map, $f(x)=1-ax^2$. The main panel shows the approximate
value of $\lambda$ for  $N=2$,
3, 4, 5 and 10. The inset shows the entropy exponent for $N=10$,
15, 20 and 25 near the accumulation point of period doubling.
In addition, the points marked by
stars are obtained by the longer, ultimately exact calculation using
kneading sequences.$^{4}$  These are not reproduced in the main panel
since they fall indistinguishably close to the calculation for $N=10$
in the parameter regime outside that drawn in more detail in the inset.
For $N=2$, which for some purposes may yield the topological entropy to
sufficient accuracy, we can write the largest eigenvalue simply as
$$
\lambda = \left\{ \matrix {
\left[ 1+\sqrt{1+4a(1-a)^2} \right] / 2 &
{\rm for} \quad  a\leq a_*, \cr
\left[ \mu + \sqrt{ \mu^2 +4a^2(a-1)(2-a) } \right] / 2(1-a) &
{\rm for} \quad a>a_*, \cr } \right.
\eqno(3)
$$
where $\mu = a(1-a)^2+a-2$ and
$$
a_* = \left[ {25 + \sqrt{621} \over 54 } \right]^{1/3}
+ \;
\left[ {25 - \sqrt{621} \over 54 } \right]^{1/3} = 1.755...
\eqno(4)
$$
is the value of $a$ such that $f$ has a period-three orbit containing
the critical point, that is, the value of the parameter where the
structure of the matrix $\bf M$ changes.

In Fig. 2, the entropy exponent is shown for the bi-modal map,
$f=x^3 - a x + b$, for $b=0.2$ and various values of
$a$. Results are shown for $N=10$ and $N=20$, and both the
approximation and bounds are displayed for each.
(Calculations for this case are also given by Bloch and Keesling$^{5}$.)
The convergence properties of the method are shown
in Fig. 3, which portrays the logarithm of the reciprocal of
the error, estimated as the difference between the upper and lower
bounds. This is equivalent to the number of digits of accuracy.
In practice, the error in the largest eigenvalue of $\bf M$ is even smaller
than this estimate.

For an example involving real data, we calculate the topological entropy
for the empirical firing map of a stimulated marine-mollusc axon. The
convergence with $N$ is shown in Fig. 4. To construct the map $f(x)$, we
fitted splines through the measurements shown in Fig. 3(d) of Hayashi
{\it et al.},$^{12}$ and the result is
shown in the inset panel of Fig. 4,
together with a set of points generated by iterating the map and adding
some random noise.  This example illustrates the ease of calculating the
topological entropy if the absolute accuracy desired is not
excessive; three or four iterations of the critical point are sufficient
to tie the entropy down to within a few percent of its true value.

Although we advocate the use of the images of the turning points for
partitions, alternative kinds of divisions could also be used. For example,
if we use backward images of the turning points, rather than forward
images, the matrix elements are again mostly either 0 or 1. The remaining
entries are all independent of one another and, when we replace the
noninteger entries by zero (or one), the largest eigenvalue of the matrix
obtained is a lower (or upper) bound on the antilog of the entropy.
This too follows from the Perron-Frobenius theorem and kneading theory,
and makes the calculation of bounds more straightforward. In fact,
the method of G\'ora and Boyarski$^{6}$ is similar to this particular case
of our algorithm. Unfortunately, computing pre-images often requires
one to solve additional equations, and the number of partitions
grows rapidly as
one iterates back. Such a partition rapidly becomes impractical.

Another partition can be generated by taking a set of periodic orbits
each of which contains a point which lies near one of the turning
points.$^{13}$ This is equivalent to
approximating $f(x)$ by a Markov map, and is accurate when the
orbits fall close to the turning points. However, a search
is required in order to find these periodic orbits, and no bounds
are given.
Finally, for experimental data,
uniform or measurement-determined partitions might be more accessible,
although they produce relatively slow convergence (as indicated
in Fig. 4 by the results of an essentially arbitrary partitioning of the
firing map).

The calculation of the topological entropy of maps or flows
with higher dimension is more complicated than in the one-dimensional
examples considered here. There are currently no effective algorithms
to compute and bound $h$ in such situations, but
property (b) above has been recently extended to
smooth diffeomorphisms of $n-$dimensional manifolds$^{14}$.
In these cases, the topological entropy determines the growth rate
of the volumes of the iterates of some submanifolds.
This suggests that we might be able to generalize
our method to higher dimension.
For certain maps and flows,
one might hope to avoid issues such as these.
For example, in work to appear with G.R.
Ierley we explain how to put the H\'enon map and some flows in $\RR^3$ into
forms suitable for producing topological transfer matrices, though the
cost in accuracy caused by
conversion to one-dimensional maps has yet to be assessed.

\medskip
We are very grateful to Alan Hoffman and John Milnor for helping
to develop some of the proofs alluded to here. We are also indebted to Roy
Adler, Karen Brucks, Steve Childress, Bruce Kitchens, Philippe Thieullen and
Alan Wolf for comments and references. This work was supported by the
A.F.O.S.R. and the S.E.R.C.

\vfil
\eject

\item{[1]} R.L. Adler, A.G. Konheim and M.H. McAndrew,
Trans. Amer. Math. Soc. {\bf 114}, 309 (1965).

\item{[2]} C. Tresser and P. Coullet, C. R. Acad. Sc. Paris A {\bf
287}, 577 (1978).

\item{[3]} M. Vishik, unpublished. See also J.M. Finn, J.D. Hanson,
I. Kan and E. Ott, Phys. Fluids B {\bf 3}, 1250 (1991).

\item{[4]} P. Collet, J.P. Crutchfield and J.-P. Eckmann, Commun. Math.
Phys. {\bf 88}, 257 (1983); L. Block, J. Keesling, S. Li and K.
Peterson, J. Stat. Phys. {\bf 55}, 929 (1989).

\item{[5]} L. Block and J. Keesling, J. Stat. Phys. {\bf 66}, 755
(1992).

\item{[6]} P. G\'ora and A. Boyarsky, Trans. Amer. Math. Soc. {\bf
323}, 39 (1991).

\item{[7]} M. Misiurewicz and W. Szlenk, Studia Mathematica {\bf 67},
45 (1980).

\item{[8]}  In one standard definition
(R. Bowen, Trans. Amer. Math. Soc. {\bf 153}, 401 (1971)),
we consider a map $f$ from the unit cube
in Euclidean space (or any compact space) to itself.  A set of points in
the cube is said to be $n-\epsilon$ separated if, for {\it each} pair of
points, $(x,y)$, in the set, with $x \ne y$, there exists an $m$ greater
than zero and less than $n$ such that $f^m(x)$and $f^m(y)$ are separated
by more than $\epsilon$.  Let ${\cal N}_{\epsilon}$ be the maximal number
of points in any set that is $n-\epsilon$ separated.  For fixed
$\epsilon$ we let $n$ go to infinity and look at the rate of increase of
the monotonic upper bound of ${\cal N}_{\epsilon}$ with
$n$.  The limit of this rate as $\epsilon \rightarrow 0$, is the
topological entropy of $f$.

\item{[9]} W.F. Gantmacher, {\sl Theory of Matrices, Vol. 1} (Chelsea
Publishing Co., 1959).

\item{[10]} See, for example,
C.S. Hsu and M.C. Kim, Phys. Rev. A {\bf 31}, 3253 (1985).


\item{[11]} J. Milnor and W. Thurston, Lecture Notes in
Mathematics {\bf 1342}, 465 (1978).

\item{[12]} H. Hayashi, S. Ishizuka, M. Ohta and K. Hirokawa, Physics
Letters {\bf 88A}, 435 (1982).

\item{[13]} L. Bloch and E. Coven, in {\sl Dynamical Systems and
Ergodic Theory}, p 237 (Polish Sci. Publishers, Warsaw, 1989).

\item{[14]} S. E. Newhouse, Ergod. Th. and Dynam. Sys. {\bf 8},
283 (1988);
Y. Yomdin, Israel J. of Math. {\bf 57}, 285 (1980), and {\bf 57},
301 (1980).

\vfill
\eject
\centerline {\bf Figure Captions}

FIG. 1. The entropy exponent for the quadratic map $f(x)=1-ax^2$.
The main panel shows the exponent calculated as the leading eigenvalue
of the topological transfer matrix with $N=2$, 3, 4, 5 and 10. In the
inset, the piece of the parameter range near the period-doubling
accumulation point is shown in greater detail for calculations with
$N=10$, 15, 20 and 25. The points marked by stars show the result of a
calculation with kneading sequences, which would not be distinguished from
the $N=10$ curve in the main panel in the parameter range outside
that shown in the inset.

\bigskip
FIG. 2. The entropy exponent for the bi-modal
cubic map $f(x)=x^3+ax+b$ with
$b=0.2$. The two panels show the approximation for the exponent
(the continuous curve) and its
two bounds (the dotted curves) for $N=10$ and $N=20$.

\bigskip
FIG. 3. The logarithm of the reciprocal of the
error in the approximation as a function
of $a$ for the cubic map ($b=0.2$), estimated as the difference between
the two bounds. Drawn in this way, it measures the number of digits of
accuracy.

\bigskip
FIG. 4. The calculation of the topological entropy for the firing map of
a sea-mollusc axon. The inset panel shows the map derived by fitting
splines to experimental data, and also a set of points derived by
iterating the map and adding random noise. The main panel shows the
convergence of the calculation of the entropy exponent with $N$ for the
partition generated by iterating the critical point,
and its bounds. A slowly converging
calculation using an essentially arbitrary partitioning of the interval
is also shown.

\end